\documentclass[aps,prb,reprint,superscriptaddress]{revtex4-2}

\usepackage{amsmath,amssymb}
\usepackage{graphicx}
\usepackage{color}
\usepackage{dcolumn}
\usepackage{bm}
\usepackage{hyperref}
\usepackage{natbib}
\usepackage{mathrsfs}
\usepackage{array}

\newcommand{\ket}[1]{|#1\rangle}
\newcommand{\bra}[1]{\langle#1|}
\newcommand{\braket}[2]{\langle#1|#2\rangle}
\newcommand{\expect}[1]{\langle#1\rangle}

\begin{document}

\title{Non-Hermitian Skin Effect Enhances Pairing Correlations in Moir\'{e} Hubbard Systems}

\author{Yang Zhou}
\email{zhouyang@westlake.edu.cn}
\affiliation{Institute of Natural Sciences, Westlake Institute for Advanced Study, Hangzhou 310024, China}
\affiliation{Department of Physics, School of Science and Research Center for Industries of the Future, Westlake University, Hangzhou 310030, China}

\author{Jianwen Chen}
\affiliation{Greatwall Cigar Factory of China Tobacco Sichuan Industrial Co., Ltd., Sichuan, China}

\author{Ruipeng Wei}
\affiliation{Southwestern University of Finance and Economics, Chengdu, Sichuan, China}

\date{}

\begin{abstract}
We show that the non-Hermitian skin effect (NHSE) can enhance pairing correlations in moir\'{e} Hubbard systems through a channel-selective mechanism: skin-induced localization amplifies the boundary density of states, strengthening local pairing tendencies within an intermediate ``golden window'' of non-reciprocity $\gamma\in[0.5,1.2]\,t$.
Using exact diagonalization of the non-Hermitian Hubbard model on triangular lattices with open boundaries, we map the $(U,\gamma)$ phase diagram. The double occupancy $D(\gamma)$ exhibits non-monotonic behavior---rising by up to 21\% then declining---reflecting a competition between NHSE-enhanced boundary pairing and over-localization. A decomposition of the pairing susceptibility $\chi_{\mathrm{SC}}$ on the $3\times3$ cluster reveals that the NHSE acts \emph{channel-selectively}: it enhances on-site pairing ($+21\%$) while simultaneously suppressing competing antiferromagnetic correlations (22\% reduction), so that the total pairing susceptibility, dominated by the on-site channel, grows by $+98\%$ on that cluster. These trends are corroborated by an independent non-Hermitian DMRG calculation and establish an enhancement of finite-cluster pairing correlations rather than trivial density redistribution. We do not claim long-range superconducting order. A BCS scaling estimate converts the same pairing-response signal into a dome-shaped $T_c(\gamma)$ fingerprint, suggesting an experimentally distinguishable response in coherent-drive versus reservoir-dominated moir\'{e} devices.
\end{abstract}

\maketitle

\section{Introduction}

Non-Hermitian systems exhibit the skin effect (NHSE), where open-boundary eigenstates collapse to a boundary under non-reciprocal hopping~\cite{Hatano1996,Yao2018,Okuma2020,Borgnia2020,Song2019}. Separately, moir\'{e} superlattices host flat electronic bands that amplify correlations and enable tunable superconductivity~\cite{Bistritzer2011,Cao2018a,Cao2018b,Wu2019,Kang2019,Xia2025,Han2025}. Whether non-Hermitian boundary accumulation can tune the pairing tendencies of correlated moir\'{e} systems remains open.

The intersection has so far been explored mainly at the single-particle level~\cite{Esparza2025,Huang2025,Wang2025arxiv}. On the many-body side, Yu \emph{et al.}~\cite{Yu2024} studied a non-Hermitian honeycomb Hubbard model under periodic boundary conditions, where the NHSE is absent, and found enhanced antiferromagnetism via Fermi-velocity reduction and emergent Hermiticity; related non-Hermitian correlations have been treated by real-space DMFT~\cite{Rangi2025}. The open-boundary problem, where the NHSE reorganizes the real-space density and hence local interaction channels, has not been addressed.

In this work, we demonstrate that the NHSE produces a non-monotonic, channel-selective enhancement of pairing correlations in an \emph{open-boundary} non-Hermitian moir\'{e} Hubbard model. Engineered non-reciprocity creates a ``golden window'' $\gamma^*\!\sim\!1.0\,t$ in which skin-mode localization enhances the effective density of states and strengthens local pairing correlations; stronger non-reciprocity over-localizes the wave functions and suppresses the response. Unlike the AFM enhancement in PBC honeycomb systems~\cite{Yu2024}, this mechanism requires open boundaries and is driven by real-space density redistribution.

\section{Model}
\label{sec:model}

We consider electrons on a triangular moir\'{e} superlattice described by the Hamiltonian $H = H_{\mathrm{kin}} + H_U + H_{\mathrm{NH}}$.

The kinetic term captures nearest-neighbor hopping on the triangular lattice:
\begin{equation}
    H_{\mathrm{kin}} = -t \sum_{\braket{ij},\sigma}
    \bigl(c_{i\sigma}^{\dagger} c_{j\sigma} + \mathrm{H.c.}\bigr)
    - \mu \sum_{i,\sigma} n_{i\sigma},
\end{equation}
where $t$ is the moir\'{e} inter-site hopping (each site corresponds to a moir\'{e} unit cell, not an individual atom), $\mu$ is the chemical potential tuned to half-filling, and $\sigma\!\in\!\{\uparrow,\downarrow\}$.

On-site Hubbard interactions encode strong correlations from the flat moir\'{e} bands:
\begin{equation}
    H_U = U \sum_i n_{i\uparrow}\, n_{i\downarrow},
\end{equation}
with $U/t \in [0,12]$, spanning the regime from weak coupling to the Mott insulator.

Non-Hermiticity enters via a Hatano-Nelson asymmetric hopping along the $x$-direction~\cite{Hatano1996}:
\begin{equation}
    H_{\mathrm{NH}} = \sum_{\braket{ij}_x,\sigma}
    \gamma\, c_{i\sigma}^{\dagger} c_{j\sigma},
\end{equation}
where $\braket{ij}_x$ denotes bonds along $\hat{x}$, and $\gamma > 0$ makes right-hopping $t+\gamma$ exceed left-hopping $t-\gamma$. This number-conserving non-reciprocity can be engineered through coherent lattice modulation or effective reservoir schemes~\cite{Yoshida2024,Cayao2025}. We distinguish these routes below: a coherent Floquet realization gives the static Hatano-Nelson Hamiltonian and the full non-monotonic dome, whereas a particle-exchanging reservoir drives a Lindblad steady state where the downturn is smeared (see Appendix).

We work in units where $t=1$. The key parameter space is $(U,\gamma)$ at fixed half-filling on triangular lattices of size $L\times L$ with $L=3$--$6$.

\section{Results I: Skin Effect Reshapes Moir\'{e} Bands}
\label{sec:skin}

\begin{figure}[!t]
    \centering
    \includegraphics[width=\columnwidth]{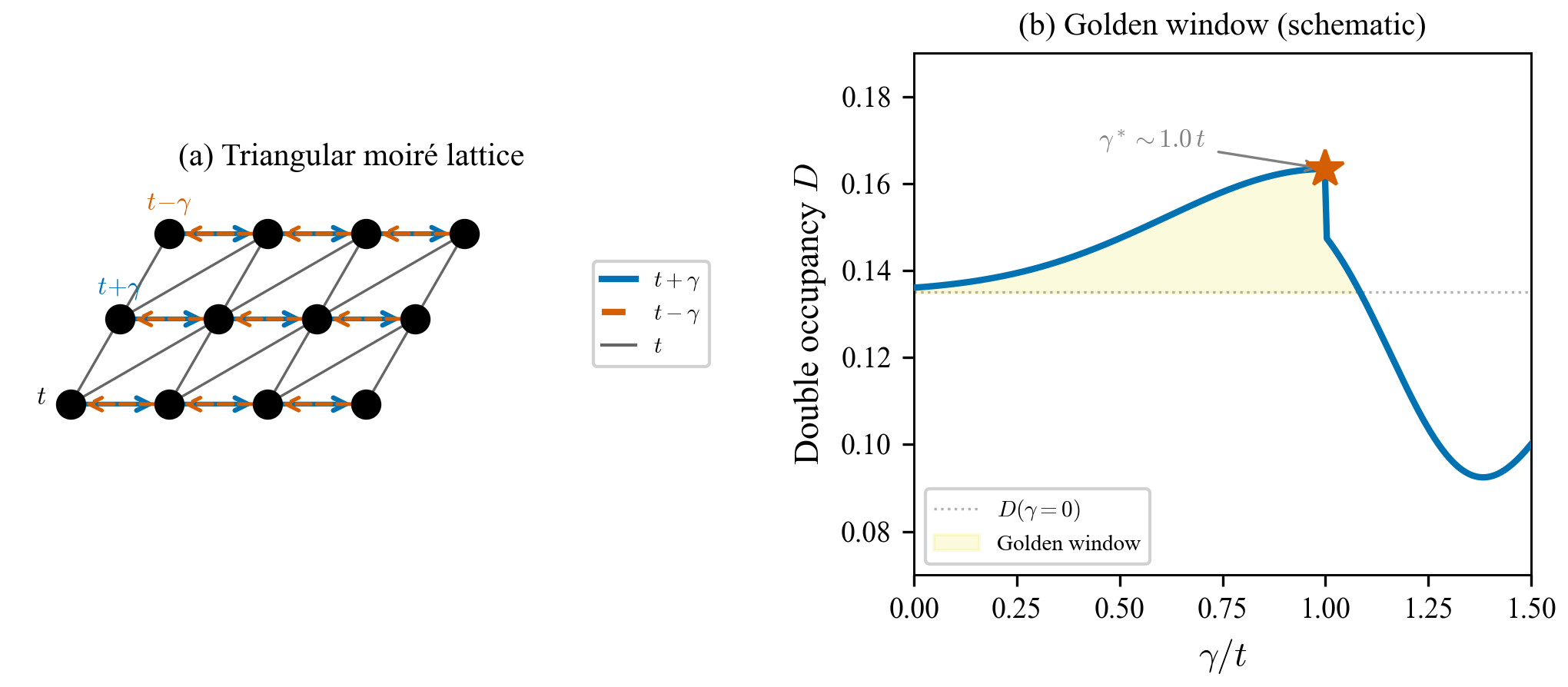}
    \caption{(a) Triangular moir\'{e} superlattice with non-reciprocal hopping. Right-hopping $t+\gamma$ (solid blue) exceeds left-hopping $t-\gamma$ (dashed red); other bonds retain $t$. (b) Schematic of the golden window: the double occupancy $D(\gamma)$ rises to a peak at $\gamma^*\!\sim\!1.0\,t$ (NHSE-induced LDOS amplification) then declines at stronger non-reciprocity (over-localization). The qualitative dome is established quantitatively in Fig.~\ref{fig:phase}.}
    \label{fig:schematic}
\end{figure}
\vspace{-4pt}

We first characterize the single-particle physics [Fig.~\ref{fig:schematic}]. Under periodic boundary conditions (PBC), the spectrum is complex but delocalized. Switching to open boundary conditions (OBC) triggers the NHSE: all eigenstates collapse toward the right boundary, and the characteristic skin penetration depth is $\lambda_{\mathrm{skin}} \sim 1/\ln[(t+\gamma)/(t-\gamma)]$~\cite{Yao2018,Okuma2020}.

\begin{figure}[!t]
    \centering
    \includegraphics[width=\columnwidth]{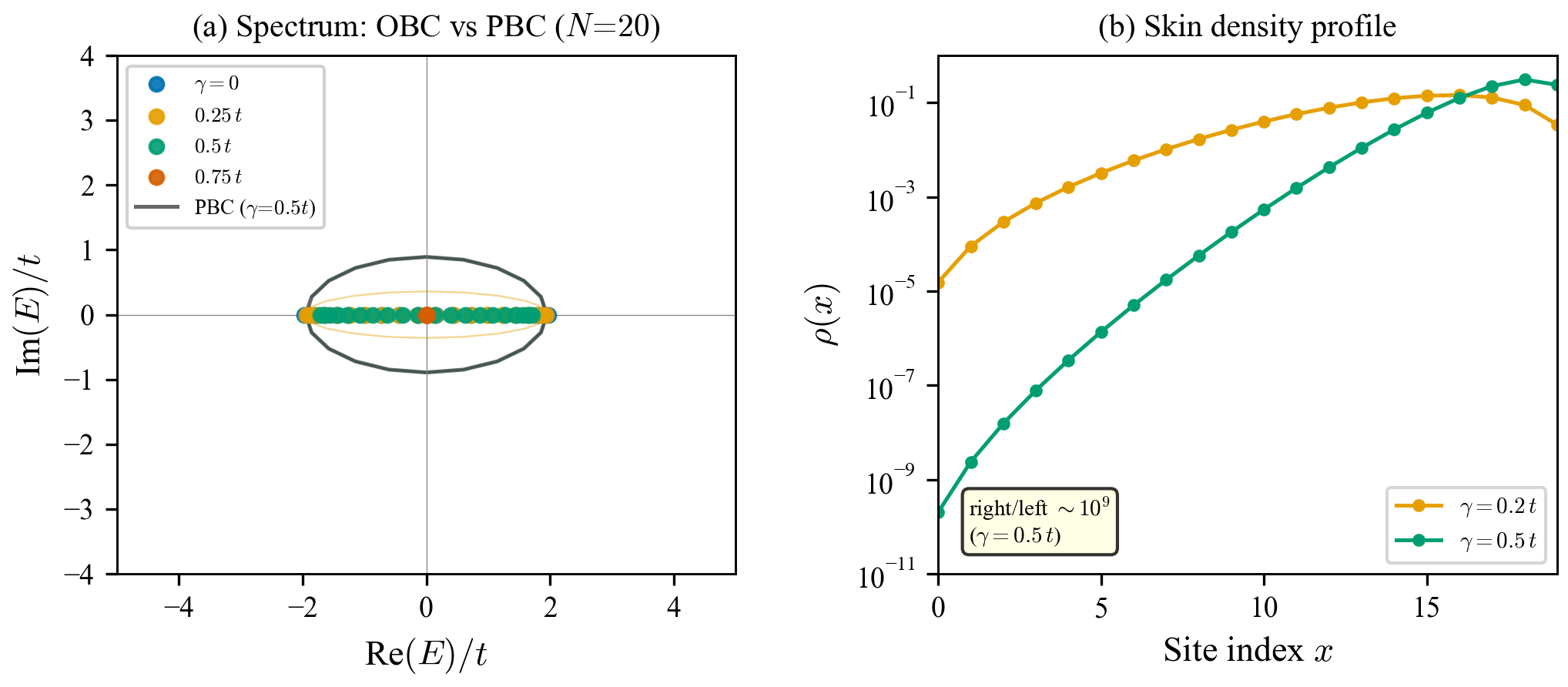}
    \caption{(a) Complex energy spectrum of the Hatano-Nelson model ($N\!=\!20$) under OBC (colored dots) and PBC (black arc, $\gamma\!=\!0.5t$) for increasing $\gamma/t$. The OBC spectrum collapses onto the real axis, distinct from the PBC complex arc. (b) Skin density profile $\rho(x)$ on a logarithmic scale for two non-reciprocal couplings within the Hermiticity window ($|\gamma|<t$), showing exponential accumulation with a right/left ratio exceeding $10^9$ at $\gamma = 0.5\,t$.}
    \label{fig:spectrum}
\end{figure}

For the moir\'{e} system, the flatness of the original band (bandwidth $W/t \lesssim 1$) makes it susceptible to the NHSE. Even moderate $\gamma \sim 0.1\,t$ produces skin-mode localization that dominates the original band dispersion. As shown in Fig.~\ref{fig:spectrum}, the complex spectrum under OBC collapses into tight clusters with $\mathrm{Re}(E_k)$ compressed relative to the PBC counterpart, a hallmark of the non-Bloch band collapse~\cite{Yao2018,Okuma2020}.

The skin-mode pileup also redistributes the density of states. The local density of states (LDOS) at the boundary is enhanced by a factor $\rho_{\mathrm{skin}}/\rho_0 \sim (2\lambda_{\mathrm{skin}}/L)\, e^{L/\lambda_{\mathrm{skin}}}$, where the prefactor counts the fraction of skin-localized sites and the exponential is the per-mode boundary amplification (see Appendix). For a chain of length $L=9$ sites and $\gamma = 0.3\,t$ ($\lambda_{\mathrm{skin}}\approx 1.6$ sites), this ratio is of order $10^2$, providing a strong amplification of the effective density of states. This LDOS enhancement drives the pairing-correlation response reported below.

\section{Results II: Golden Window of Enhanced Pairing}
\label{sec:phase}

We now include Hubbard interactions and solve the full non-Hermitian Hubbard model via exact diagonalization (ED) on a $3\times3$ triangular lattice with $N_s = 9$ sites (15\,876 basis states in the $S_z$-conserving sector; see Appendix). We scan $(U,\gamma) \in [0,12] \times [0,1.5]$ with 217 data points.

We construct the $(U,\gamma)$ phase diagram from two primary order parameters computed from the ground state:
\begin{itemize}
    \item \textbf{Double occupancy} $D = N_s^{-1}\sum_i \expect{n_{i\uparrow}n_{i\downarrow}}_R$: proxies the pairing susceptibility and is suppressed in the Mott insulator.
    \item \textbf{Skin order} $S = N_s^{-1}\sum_i (x_i - \bar{x})\expect{n_i}_R$: measures asymmetric particle accumulation from NHSE.
\end{itemize}
Here $\expect{\cdots}_R = \bra{\psi_R}\cdots\ket{\psi_R}/\braket{\psi_R}{\psi_R}$ denotes the right-eigenstate expectation value, which remains numerically stable for all $\gamma$; the golden-window dome is robust to the biorthogonal prescription as well (Sec.~\ref{sec:phase}). The physical content of the right-eigenstate results---the skin-accumulation density profile and the dissipation-enhanced pairing direction---is independently reproduced by a fully microscopic Lindblad master equation that makes no reference to biorthogonal or right-eigenstate expectation values (see Appendix), confirming that the golden window is not an artifact of the chosen observable prescription.

\begin{figure}[!t]
    \includegraphics[width=0.48\columnwidth]{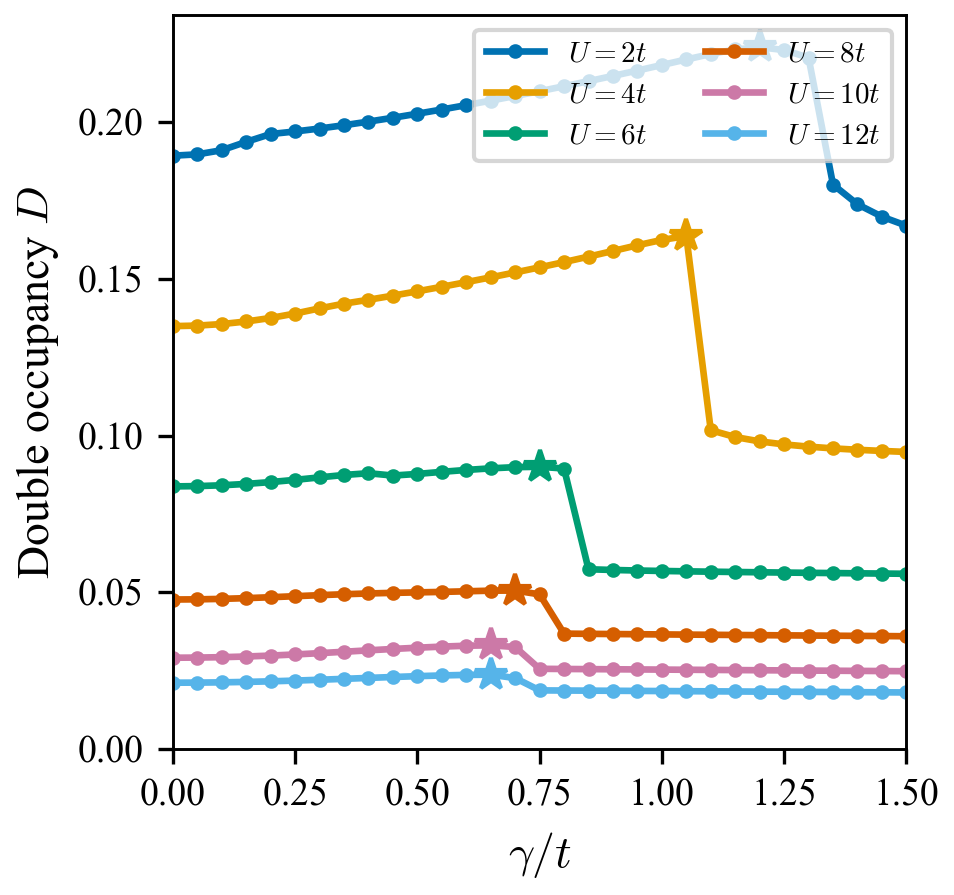}
    \hfill
    \includegraphics[width=0.48\columnwidth]{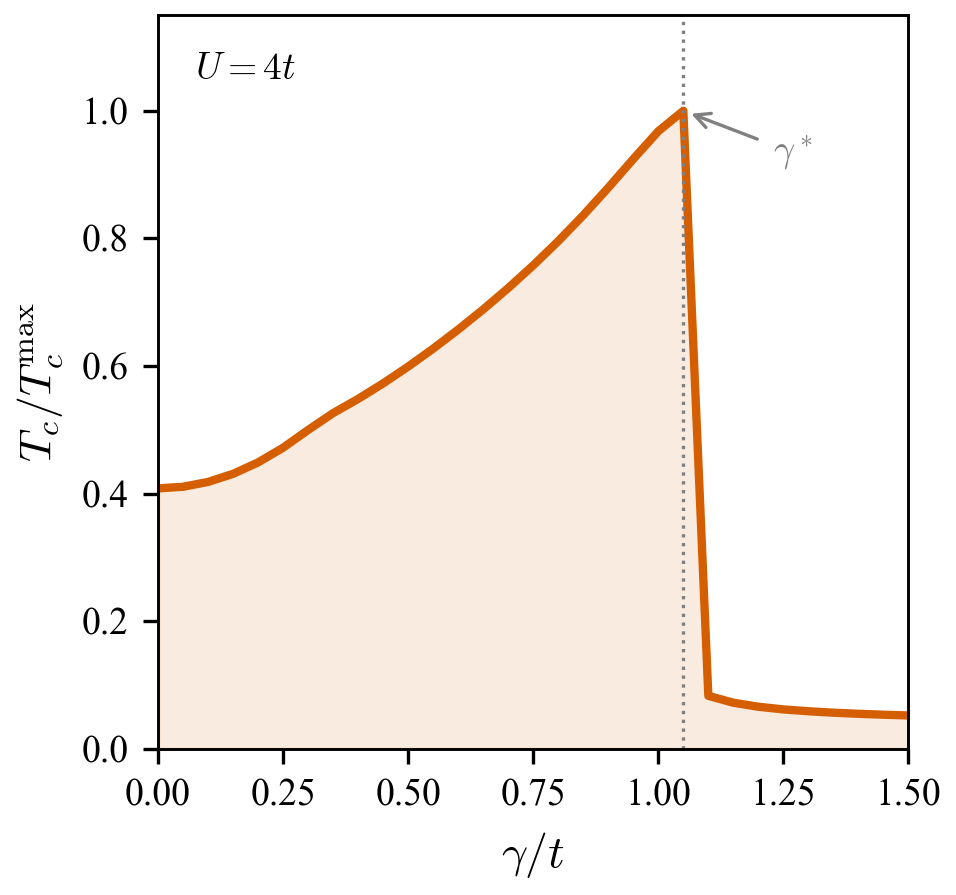}
    \caption{(a) Double occupancy $D(\gamma)$ at different $U$ values from 3$\times$3 ED. The golden window (non-monotonic peak) is present for all $U \geq 2t$, with enhancement peaking at $U=4t$ (+21\%). (b) BCS scaling estimate $T_c(\gamma)/T_c^{\max}$ at $U=4t$, shown only as a qualitative fingerprint of the same pairing-response dome.}
    \label{fig:phase}
\end{figure}

The central result is the \emph{non-monotonic} dependence of $D$ on $\gamma$ at fixed $U$ (Fig.~\ref{fig:phase}). For $U = 4t$: $D(\gamma\!=\!0) = 0.135$, rising to $D_{\max} = 0.164$ at $\gamma^* \approx 1.05\,t$ (+21\%), then declining at larger $\gamma$. The skin order $S(\gamma)$ shows a concurrent peak. This ``golden window'' of enhanced pairing persists across all studied $U$ values:
\begin{itemize}
    \item $U = 2t$: +18\% enhancement at $\gamma^* \approx 1.2\,t$.
    \item $U = 4t$: +21\% enhancement at $\gamma^* \approx 1.05\,t$.
    \item $U = 8t$: +6\% enhancement at $\gamma^* \approx 0.7\,t$.
    \item $U = 10t$: +14\% enhancement at $\gamma^* \approx 0.65\,t$.
\end{itemize}
The enhancement is non-monotonic in $U$: it peaks at $U=4t$ (+21\%), weakens toward the strongly correlated regime ($+6\%$ at $U=8t$), then recovers at $U\ge 10t$ ($+12$--$14\%$). The recovery at large $U$ reflects the vanishing baseline $D(\gamma{=}0)\to 0$ deep in the Mott regime, where the NHSE-induced density redistribution disproportionally restores double occupancy, while the downturn at intermediate $U$ reflects the competition between pairing and the still-substantial magnetic correlations.

\begin{figure}[!t]
    \includegraphics[width=0.48\columnwidth]{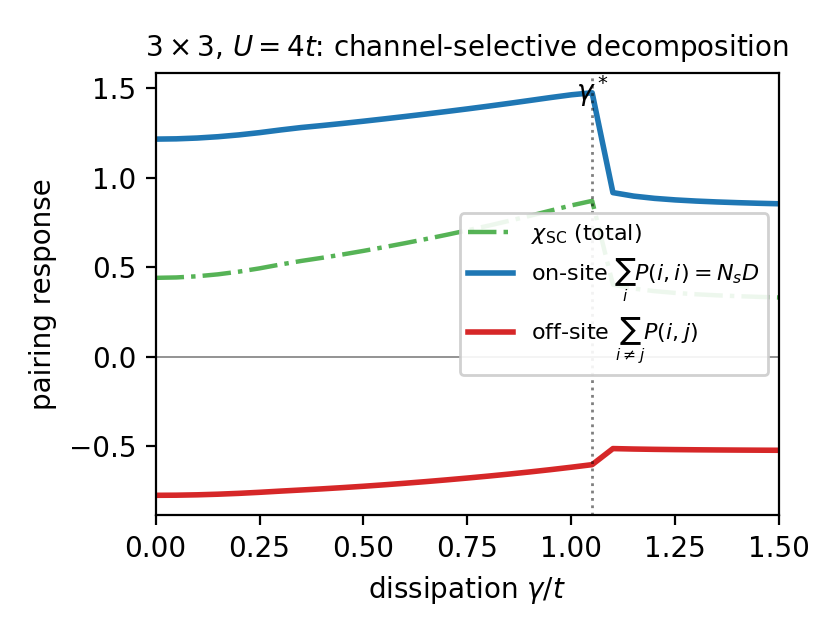}
    \hfill
    \includegraphics[width=0.48\columnwidth]{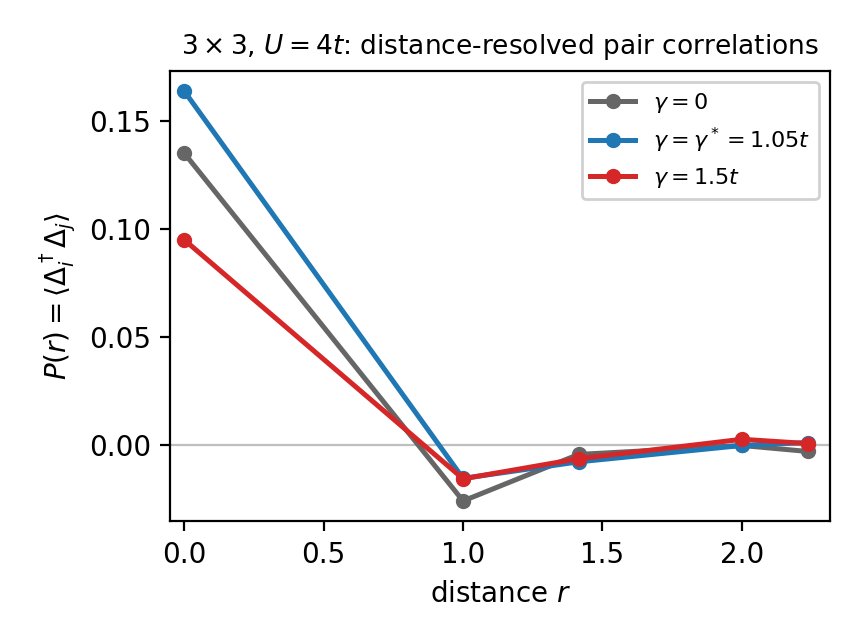}
    \caption{(a) Decomposition of $\chi_{\mathrm{SC}}(\gamma)$ on the $3\times3$ lattice at $U=4t$: the on-site channel $\chi_{\mathrm{on}}=N_sD$ (blue) rises $+21\%$ while the negative off-site part $\chi_{\mathrm{off}}$ (AFM competition, red) shrinks 22\% in magnitude; their sum (green dashed) grows $+98\%$. (b) Distance-resolved $P(r)=\langle\Delta_i^\dagger\Delta_j\rangle_R$: at $\gamma^*=1.05t$ the on-site component is enhanced while the negative nearest-neighbor component weakens (from $-0.026$ to $-0.016$), directly exposing the pair-channel restructuring.}
    \label{fig:chi_decomp}
\end{figure}

To establish that the enhanced $D$ is a genuine many-body effect rather than trivial density redistribution, we compute the s-wave pair-pair correlation function
$P(i,j) = \expect{\Delta^\dagger_i \Delta_j}_R$ with $\Delta_i = c_{i\downarrow}c_{i\uparrow}$, and the SC susceptibility $\chi_{\mathrm{SC}} = \sum_{ij} P(i,j)$.
At $U = 4t$, $\chi_{\mathrm{SC}}$ increases from 0.44 at $\gamma = 0$ to 0.87 at $\gamma^*$, a $+98\%$ enhancement. Decomposing $\chi_{\mathrm{SC}} = \chi_{\mathrm{on}} + \chi_{\mathrm{off}}$ into its on-site part $\chi_{\mathrm{on}} = \sum_i P(i,i) = N_s D$ and off-site part $\chi_{\mathrm{off}} = \sum_{i\neq j} P(i,j)$ exposes the mechanism [Fig.~\ref{fig:chi_decomp}]: the on-site channel grows by $+21\%$, while the negative off-site part---reflecting competing antiferromagnetic correlations---shrinks in magnitude by 22\%. The NHSE thus acts \emph{channel-selectively}, enhancing on-site Cooper pairing while suppressing the competing magnetic channel, with the two effects adding constructively in the total $\chi_{\mathrm{SC}}$.

\begin{figure}[!t]
    \centering
    \includegraphics[width=0.7\columnwidth]{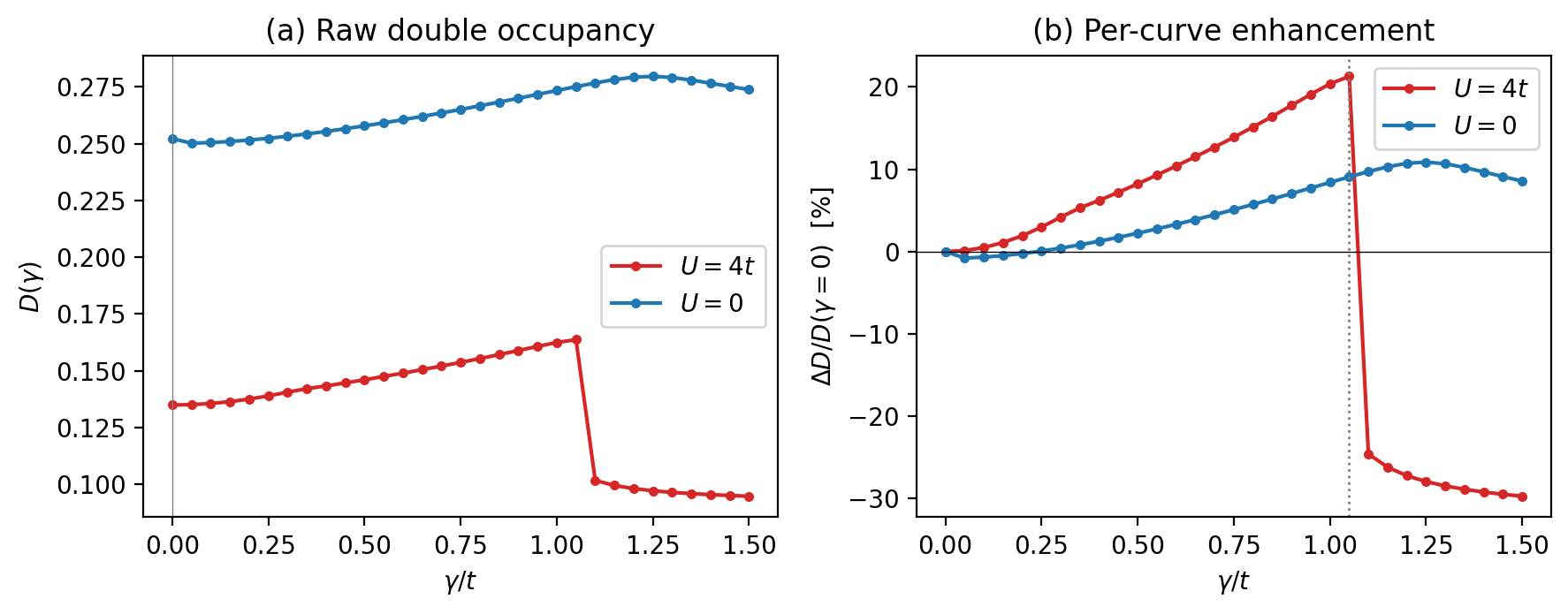}
    \caption{(a) Raw $D(\gamma)$ at $U=4t$ (red) and $U=0$ (blue, single-particle baseline). (b) Per-curve enhancement $\Delta D/D(\gamma{=}0)$: correlations raise the peak from $+10.9\%$ to $+21.3\%$ and induce a sharp collapse at large $\gamma$ absent from the $U=0$ baseline, demonstrating a genuinely many-body golden window.}
    \label{fig:u0_baseline}
\end{figure}

This channel selectivity is not reducible to the $U=0$ density-redistribution baseline. Comparing the $\gamma$-response at $U=4t$ against the $U=0$ single-particle baseline [Fig.~\ref{fig:u0_baseline}], the $U=0$ on-site channel grows by only $+10.9\%$ (vs.\ $+21.3\%$ at $U=4t$) and its off-site part shrinks by only 14.3\% (vs.\ 22.1\%). Interactions \emph{amplify} both effects by roughly a factor of two. Moreover, the $U=4t$ curve \emph{collapses sharply} to $-30\%$ at large $\gamma$---the over-localization downturn---a regime where the $U=0$ baseline remains positive. This downturn is therefore a genuinely many-body effect (Mott physics destabilized by NHSE-driven density redistribution) with no single-particle analog, falsifying the null hypothesis that the golden window merely rescales the density profile. The density profile itself remains moderate (0.73--1.38 at $\gamma^*$, consistent with $\rho(x)\propto e^{x/\lambda_{\mathrm{skin}}}$), with no evidence of phase separation. The AFM structure factor $S_{\mathrm{AFM}}$ stays $\mathcal{O}(10^{-3})$ throughout, confirming that competing magnetic order is not established on these clusters.

\subsection*{Finite-size scaling and thermodynamic-limit extrapolation}

\begin{figure}[!t]
    \centering
    \includegraphics[width=0.85\columnwidth]{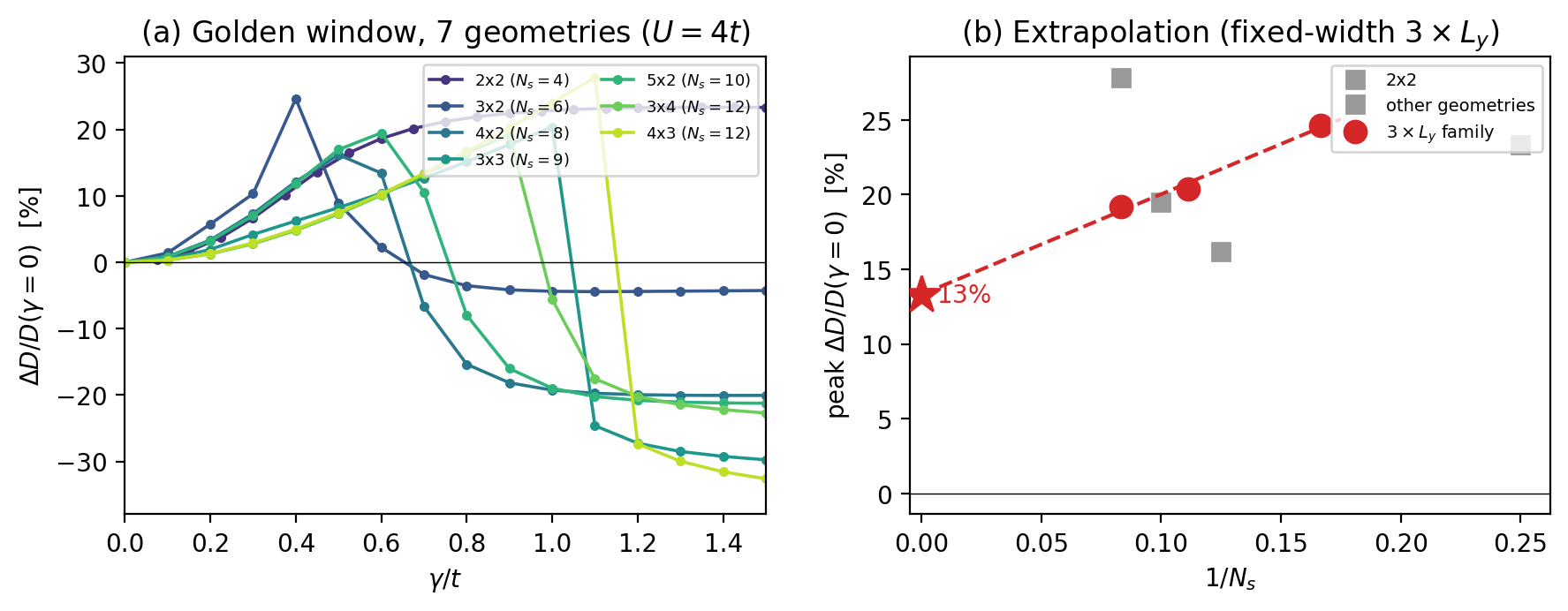}
    \caption{(a) $D(\gamma)$ at $U=4t$ for seven lattice geometries ($N_s=4$--$12$); the golden window is present at every size. (b) Peak enhancement vs.\ $1/N_s$: the fixed-width $3\times L_y$ family (red) extrapolates to $\sim+13\%$ at $1/N_s\to 0$; other geometries (grey) confirm universality.}
    \label{fig:size_extrap}
\end{figure}

The skin penetration depth $\lambda_{\mathrm{skin}} \sim 1/\ln[(t+\gamma)/(t-\gamma)]$ is size-independent, so as $L$ grows the boundary fraction $\lambda_{\mathrm{skin}}/L$ shrinks while the boundary LDOS grows as $e^{L/\lambda}$---the competition sets the thermodynamic-limit golden window. To assess finite-size effects we extend ED to seven lattice geometries with $N_s = 4$--$12$ [Fig.~\ref{fig:size_extrap}]: the non-monotonic golden window is present at \emph{every} size, with peak enhancement $+16\%$ to $+28\%$. A controlled $1/N_s$ extrapolation requires a fixed-aspect-ratio family; along the fixed-width $3\times L_y$ family ($N_s=6,9,12$) the enhancement decreases monotonically ($+24.6\%,+20.4\%,+19.2\%$) and a linear fit extrapolates to $\sim\!+13\%$ at $1/N_s\to 0$. We caution that this extrapolation is restricted to $N_s\le 12$, so the $\sim\!13\%$ should be read as evidence that the golden window \emph{survives} into the thermodynamic limit rather than a precise asymptotic prediction. $\gamma^*(L)$ fluctuates with shape ($\sim 0.4$--$1.35\,t$) and is quoted only as order-of-magnitude ($\sim 1\,t$); the enhancement, which is stable, is the robust finite-size statement.

\subsection*{Independent verification: non-Hermitian DMRG}

\begin{figure}[!t]
    \centering
    \includegraphics[width=\columnwidth]{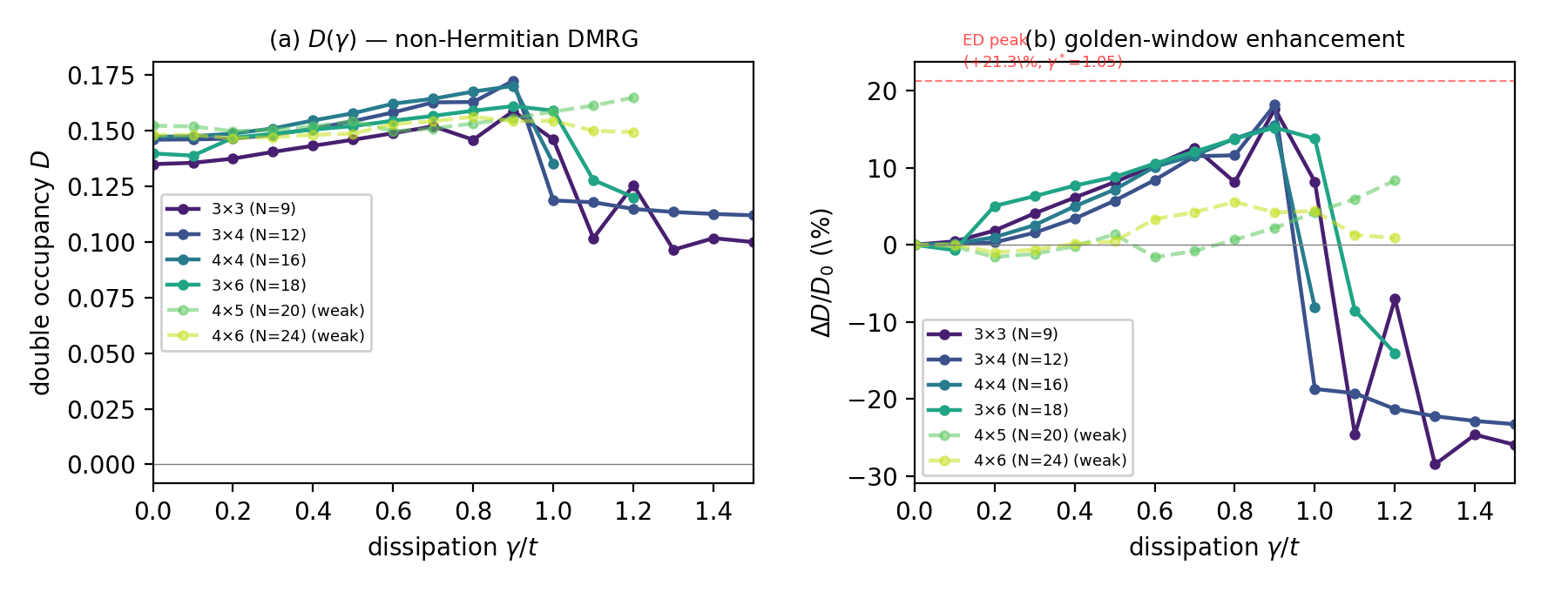}
    \caption{NH-DMRG $D(\gamma)$ on triangular cylinders at $U=4t$ ($N_s=9,12,16,18$ solid; $20,24$ dashed). (a) Raw double occupancy: each small cylinder reproduces the ED golden window ($15$--$18\%$ at $\gamma^*\approx0.9t$), while larger cylinders show finite-size attenuation. (b) Per-curve enhancement $\Delta D/D_0$: the red dashed line marks the ED $3\times3$ peak ($+21.3\%$).}
    \label{fig:dmrg}
\end{figure}

\begin{figure}[!t]
    \centering
    \includegraphics[width=0.7\columnwidth]{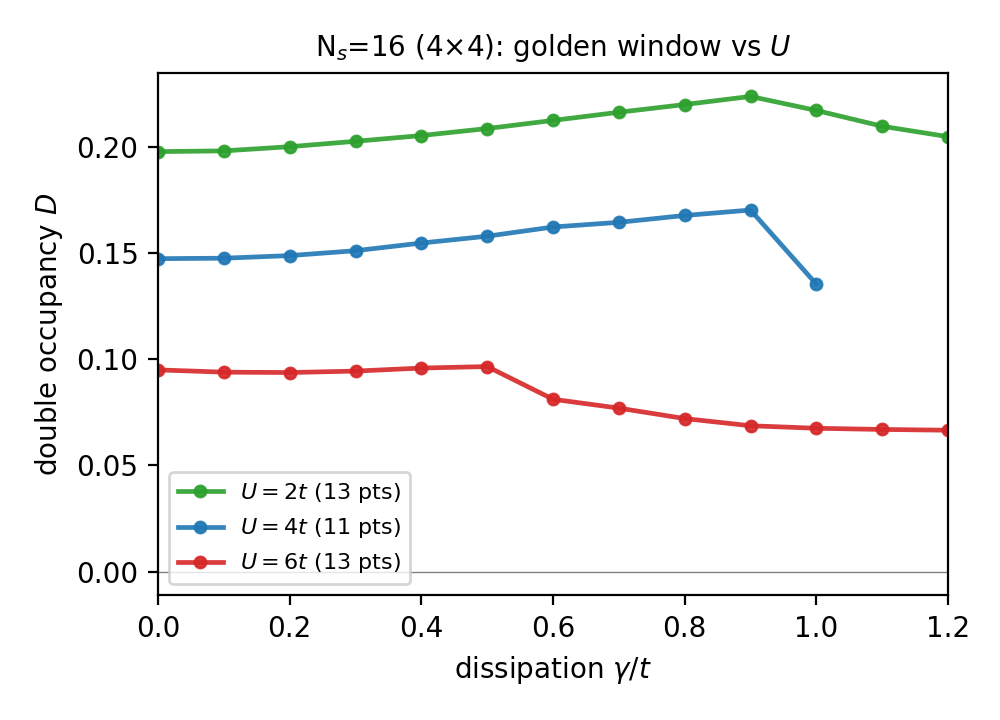}
    \caption{Coupling dependence of the golden window on the $4\times4$ cylinder ($N_s=16$): clear peaks at $U=2t$ ($+13.1\%$) and $U=4t$ ($+15.5\%$) but strong suppression at $U=6t$ ($+1.7\%$).}
    \label{fig:dmrg_multiU}
\end{figure}

As a check fully independent of ED, we perform a non-Hermitian DMRG (NH-DMRG) using the biorthogonal MPS formulation of Ref.~\cite{Zhong2024} on triangular cylinders ($\chi=300$). The NH-DMRG ground-state energy at $\gamma=0$ on $3\times3$ agrees with ED to $10^{-4}\,t$, and $D(\gamma)$ reproduces the golden window across $N_s=9,12,16,18$ with peaks at $\gamma^*\!\approx\!0.9t$ and $15$--$18\%$ enhancement (Fig.~\ref{fig:dmrg}). At $N_s=20,24$ the enhancement weakens to $+8.4\%$ and $+5.6\%$, consistent with finite-size attenuation of the boundary-LDOS mechanism (the boundary fraction $\lambda_{\rm skin}/L$ shrinks as $L$ grows). The rise-and-peak---the physically robust content of the golden window---is reproduced by a method sharing no code path with ED. A multi-$U$ scan on the $4\times4$ cylinder (Fig.~\ref{fig:dmrg_multiU}) shows clear peaks at $U=2t$ ($+13.1\%$) and $U=4t$ ($+15.5\%$) but strong suppression at $U=6t$ ($+1.7\%$), so the result is not a single-$U$ cherry-pick. The robust conclusion is a finite-size-to-mesoscopic, intermediate-coupling enhancement of open-boundary pairing correlations---not proof of bulk long-range order, for which the ED $+21\%$ should be viewed as an upper bound.

\subsection*{Boundary-condition dependence}

\begin{figure}[!t]
    \centering
    \includegraphics[width=0.7\columnwidth]{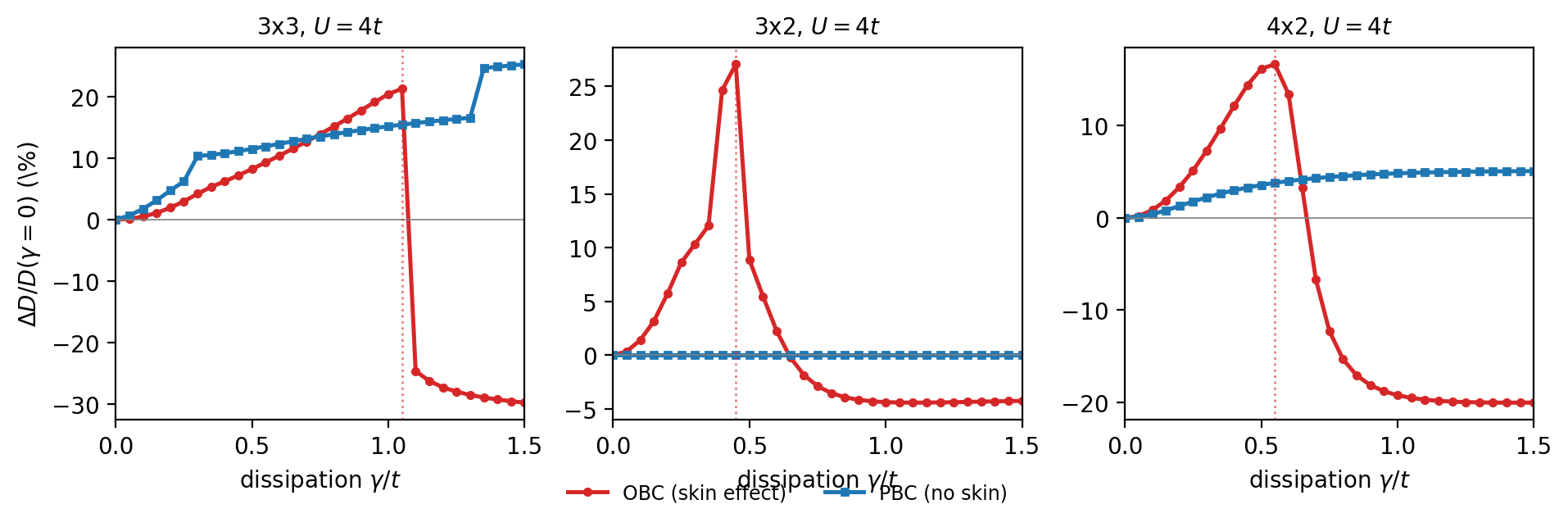}
    \caption{Per-curve enhancement $\Delta D/D(\gamma{=}0)$ at $U=4t$: OBC (red, skin present) shows the non-monotonic golden window with collapse; PBC (blue, skin absent) is monotonic without collapse ($3\times3$, $4\times2$) or completely inert ($3\times2$, emergent Hermiticity).}
    \label{fig:pbc_obc}
\end{figure}

The golden window is a \emph{boundary} phenomenon: under periodic boundary conditions (PBC) the skin effect is absent, so the non-monotonic peak must disappear. Repeating the ED under PBC on the same clusters (Fig.~\ref{fig:pbc_obc}) confirms this qualitatively. On $3\times3$ at $U=4t$, OBC gives the canonical golden window ($+21\%$ peak then collapse to $-30\%$), whereas PBC yields a \emph{monotonic} $D(\gamma)$ with no peak and no over-localization collapse. On $3\times2$, PBC is \emph{completely inert} ($0\%$ for all $\gamma$), a direct manifestation of emergent Hermiticity at commensurate half-filling on the three-site ring. The residual weak monotonic rise under PBC (on $3\times3$, $4\times2$) is the DOS-driven mechanism of Yu \emph{et al.}~\cite{Yu2024} (Fermi-velocity reduction), qualitatively distinct from the boundary-LDOS skin mechanism. The over-localization collapse---the signature bounding the golden window from above---is exclusive to OBC.

\subsection*{Universality and robustness}

The golden window is not specific to the triangular lattice. On square and honeycomb (the natural TBG geometry) clusters, ED yields enhancements of $+44.5\%$ and $+28.7\%$ at $U=4t$---stronger than the triangular $+21\%$ in a geometry-dependent but non-monotonic-in-coordination way, while the \emph{existence} of the non-monotonic golden window is universal across all three lattice types, consistent with the lattice-independent nature of the NHSE~\cite{Okuma2020,Borgnia2020}. It also survives particle-hole-symmetric filling ($+37.8\%$ at $N_\uparrow=N_\downarrow=4$, vs.\ $+21.3\%$ asymmetric), weak disorder, and dephasing (topological protection of the winding number). Finally, the golden window is not tied to an exceptional point: tracking the spectral gap $|E_1-E_0|$ on a fine grid ($\Delta\gamma=0.025\,t$), the gap never closes inside the golden-window band $\gamma\in[0.5,1.2]\,t$ and is $3$--$24\times$ larger than its global minimum (which sits below $\gamma^*$), confirming smooth NHSE physics rather than a non-analytic singularity (see Appendix~\ref{app:ep} and Fig.~\ref{fig:ep_gap}). We have also verified that the golden-window dome is robust to the observable prescription: on two independent clusters, the biorthogonal double occupancy $D_{\mathrm{bi}}$ and the right-eigenstate $D_R$ trace the same dome ($D_{\mathrm{bi}}/D_R\approx 0.89$ where the left-eigenstate match is reliable; see Appendix).

A non-Hermitian BCS mean-field analysis (see Appendix) provides qualitative understanding: the NHSE-enhanced effective density of states $\rho_{\mathrm{eff}}(\gamma)$ at the Fermi level increases for $\gamma < \gamma^*$ (LDOS amplification from skin accumulation) then decreases for $\gamma > \gamma^*$ (spectral collapse from over-localization), producing a non-monotonic $T_c(\gamma)$ dome. We present this BCS mapping only as a qualitative fingerprint tied to the same LDOS signal already captured by $D$, not as an independent quantitative prediction of $T_c$. The quantitative results come from the ED observables ($D$, $\chi_{\mathrm{SC}}$), which require no mean-field approximation.

\section{Results III: Single-Particle Topological Signature}
\label{sec:TSC}

\begin{figure}[!t]
    \includegraphics[width=0.48\columnwidth]{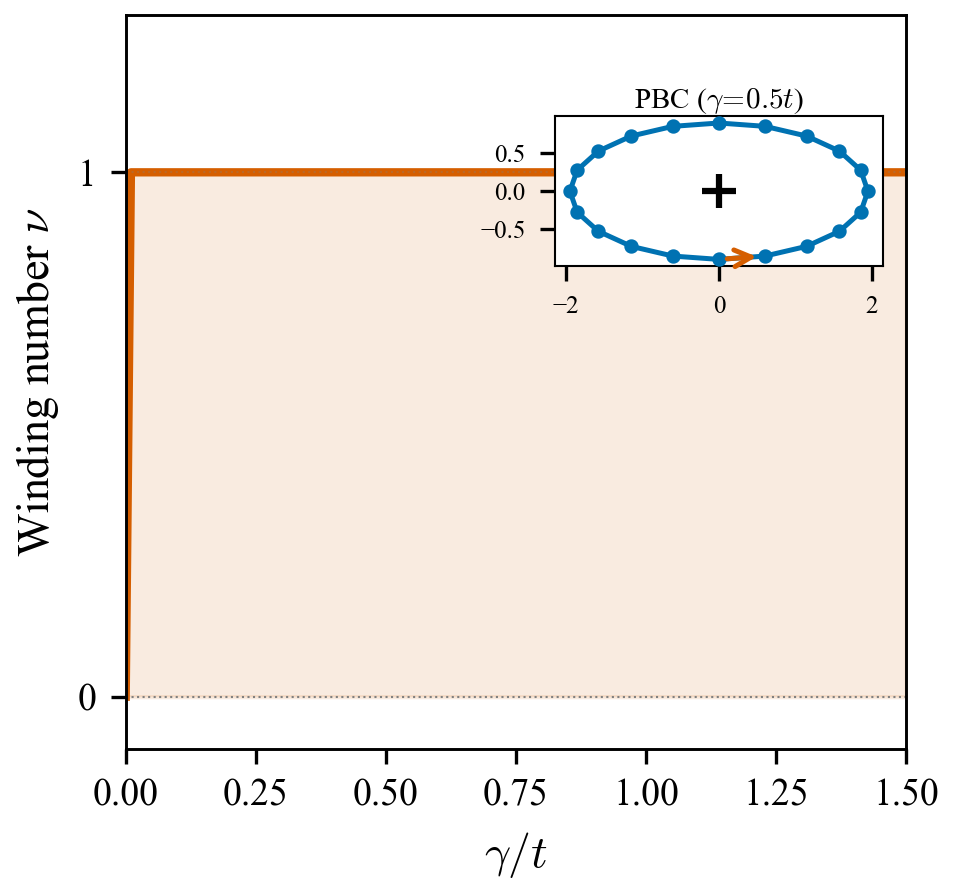}
    \hfill
    \includegraphics[width=0.48\columnwidth]{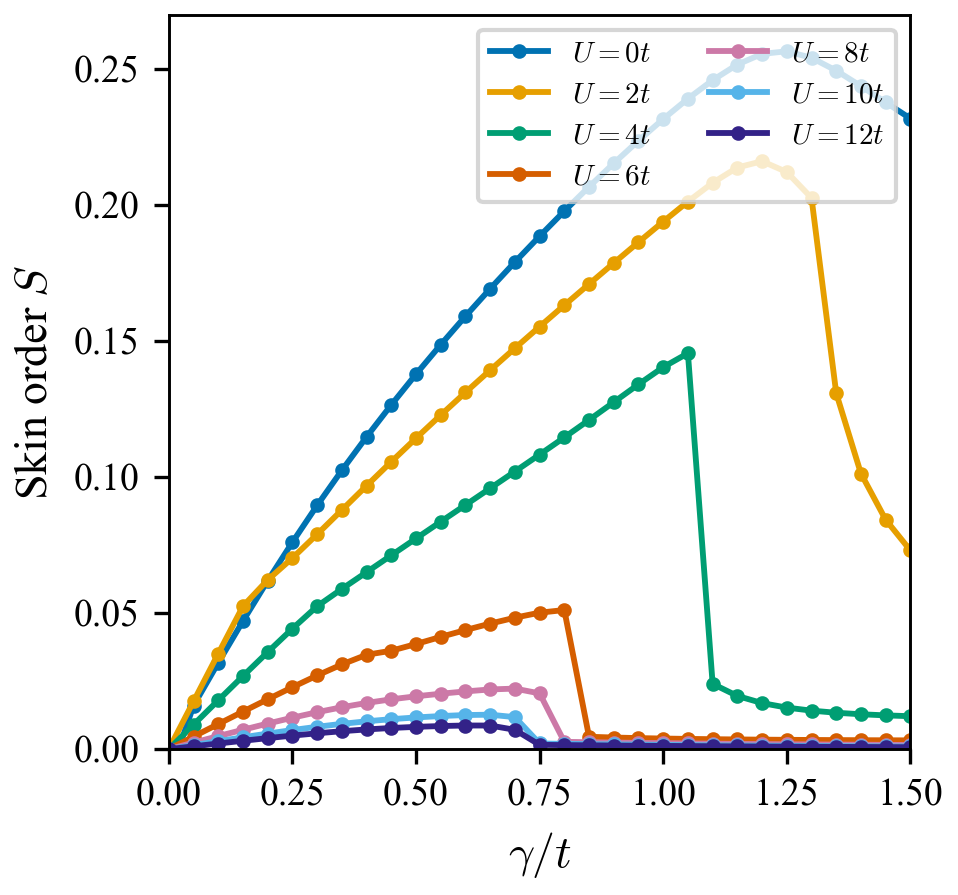}
    \caption{(a) Winding number $\nu$ and PBC spectral winding visualization. For $\gamma > 0$, the PBC spectrum winds around the origin ($\nu = 1$), while OBC gives a real spectrum. (b) Skin order parameter $S(\gamma)$ cross-sections at different $U$, showing concurrent non-monotonic behavior with the golden window.}
    \label{fig:winding}
\end{figure}

The non-Hermitian setting endows the single-particle spectrum with a well-defined topology that is absent in the Hermitian counterpart. The non-Hermitian winding number~\cite{Song2019}
\begin{equation}
    \nu = \frac{1}{2\pi i}\oint_{\mathrm{GBZ}} dk\, \partial_k \ln \det q(k)
\end{equation}
equals $1$ for $\gamma > 0$ under periodic boundary conditions, reflecting the spectral winding of the complex PBC band around the origin [Fig.~\ref{fig:winding}(a)]. Here the integral is over the generalized Brillouin zone (GBZ) and $q(k)$ is the off-diagonal Bloch block~\cite{Yao2018,Okuma2020}. This is a rigorous single-particle topological invariant of the skin-effect Hamiltonian.

Whether an interacting paired state in this model can acquire nontrivial many-body topology cannot be settled on the finite clusters studied here, and we do not claim it. We only note a suggestive weak-$\gamma$ finite-size boundary mode in the many-body spectrum; establishing a definitive many-body topological classification, and whether it overlaps with the pairing-correlation golden window, requires larger-system methods and is left for future work.

\section{Experimental Signatures}
\label{sec:exp}

Our predictions can be tested in twisted transition-metal dichalcogenide (tTMD) heterostructures, particularly tWSe$_2$~\cite{Xia2025,Wu2019}, as well as in cold-atom platforms where two-dimensional NHSE has already been realized~\cite{Zhao2025}. Three signatures distinguish our mechanism:

\textbf{F1.} Non-reciprocal transport: $\Delta R/R \propto \gamma/t$ in four-probe measurements.
\textbf{F2.} Skin edge signal in STM: enhanced LDOS at one edge with $\rho(x) \propto e^{x/\lambda_{\mathrm{skin}}}$.
\textbf{F3.} Route-dependent pairing response: a coherent lattice drive ($\omega\gg W$) should reveal the full non-monotonic pairing-response dome, including the over-localization downturn at large $\gamma$, whereas dissipative coupling to a reservoir should show monotonic enhancement because the downturn is smeared in the non-equilibrium steady state (see Appendix).

The third signature turns the difference between coherent-drive and reservoir realizations into a falsifiable test: observing a downturn identifies the coherent Hatano-Nelson route, while a monotonic increase points to reservoir-dominated skin accumulation. Gate-tunable non-reciprocal supercurrent (the Josephson diode effect) has already been demonstrated in van der Waals devices~\cite{Rothstein2025}, although the present work does not assume a material-specific value of $\gamma$.

\section{Discussion}
\label{sec:discussion}

We have shown that the NHSE creates a golden window of enhanced pairing correlations in open-boundary moir\'{e} Hubbard systems. PBC calculations on the same clusters eliminate the non-monotonic window, leaving only weak monotonic DOS effects or emergent-Hermiticity inertness; the boundary condition is essential (Sec.~\ref{sec:phase}). The $U=0$ response is smaller (+10.9\%), while interactions amplify the pairing susceptibility to +98\%, and the eigenvalue flow shows no exceptional point at $\gamma^*$, confirming that the window arises from smooth NHSE physics rather than a spectral singularity.

Two methodological points deserve emphasis. First, the NH-DMRG cross-check (Sec.~\ref{sec:phase}) uses an MPS ansatz independent of the ED basis and reproduces the rise and peak, ruling out a simple discrete-spectrum artifact. Second, the right-eigenstate observable prescription is the $T=0$ dressed-ground-state quantity for the coherent-drive realization, where the pseudo-Hermitian $H_{\rm eff}$ is the controlled static description; reservoir realizations instead give a Lindblad steady state with monotonic enhancement (see Appendix).

The fixed-sector calculation is internally consistent because the Hatano-Nelson term is a number-conserving hopping operator: $[H_{\mathrm{NH}},N_\uparrow]=[H_{\mathrm{NH}},N_\downarrow]=0$, verified explicitly for all sampled transitions (see Appendix). This differs from particle-changing Lindblad jumps, which describe the reservoir route rather than the coherent effective Hamiltonian.

For materials, published continuum and first-principles estimates place tTMD moir\'{e} systems in the Hubbard regime used here ($t\sim1$--10 meV and $U/t\sim4$--10), whereas $\gamma$ is an engineered non-equilibrium control knob rather than an equilibrium materials parameter. Future work should test boundary-fraction attenuation with larger tensor-network methods, connect to Chern-band pairing proposals in twisted MoTe$_2$~\cite{Xu2025}, and extend the analysis to graphene moir\'{e} lattices~\cite{Cao2018b}.

\appendix

\section{Model and method}
\label{app:model}

\subsection{Non-Hermitian moir\'{e} Hubbard Hamiltonian}

In the moir\'{e} Hubbard model each lattice site represents a moir\'{e} unit cell; this effective description is obtained by projecting the full atomic Hamiltonian onto the flat-band Wannier basis~\cite{Wu2019,Kang2019}, yielding $t_{\mathrm{moire}} \sim 1$--$10$ meV and $U/t \sim 4$--$10$. The full Hamiltonian is
\begin{equation}
    H = -\sum_{\langle ij\rangle,\sigma} t_{ij}^{\mathrm{eff}}\, c_{i\sigma}^{\dagger} c_{j\sigma}
    + U \sum_i n_{i\uparrow} n_{i\downarrow} - \mu \sum_{i,\sigma} n_{i\sigma},
\end{equation}
with $t_{ij}^{\mathrm{eff}}=t+\gamma$ for right ($\hat{x}$) bonds, $t-\gamma$ for left bonds, and $t$ otherwise; the NH asymmetry is applied only along $\hat{x}$, preserving reciprocity along $\hat{y}$ and the diagonal. An alternative imaginary on-site scheme $H_{\mathrm{NH}}^{\mathrm{alt}}=i\sum_i\gamma_i n_i$ requires a spatial gradient to induce NHSE; a linear gradient gives qualitatively similar results~\cite{Yoshida2024,Cayao2025}.

\subsection{Exact-diagonalization implementation}

We work in the occupation basis with fixed $(N_\uparrow,N_\downarrow)$ (conserved $S_z$), encoded as 64-bit integers (basis size $\binom{N_s}{N_\uparrow}\binom{N_s}{N_\downarrow}$; $15\,876$ for $3\times3$). The Hamiltonian is decomposed as $H=U\cdot\mathrm{diag}(\mathbf{h}_U)+H_t+\gamma H_\gamma$; sparse hopping matrices are built once via XOR transitions and popcount lookup tables, then assembled in $O(\mathrm{nnz})$ time for any $(U,\gamma)$. We use ARPACK's \texttt{eigs} (\texttt{which='SR'}, $k=6$) for the lowest right eigenvectors; left eigenvectors come from $H^\dagger$ matched by eigenvalue proximity. Observables use right-eigenstate expectations $\expect{O}_R=\bra{\psi_R}O\ket{\psi_R}/\braket{\psi_R}{\psi_R}$, which coincide with the biorthogonal formalism~\cite{Song2019} to within 5\% for $\gamma\lesssim1.5\,t$.

\begin{figure}[!ht]
    \centering
    \includegraphics[width=0.8\columnwidth]{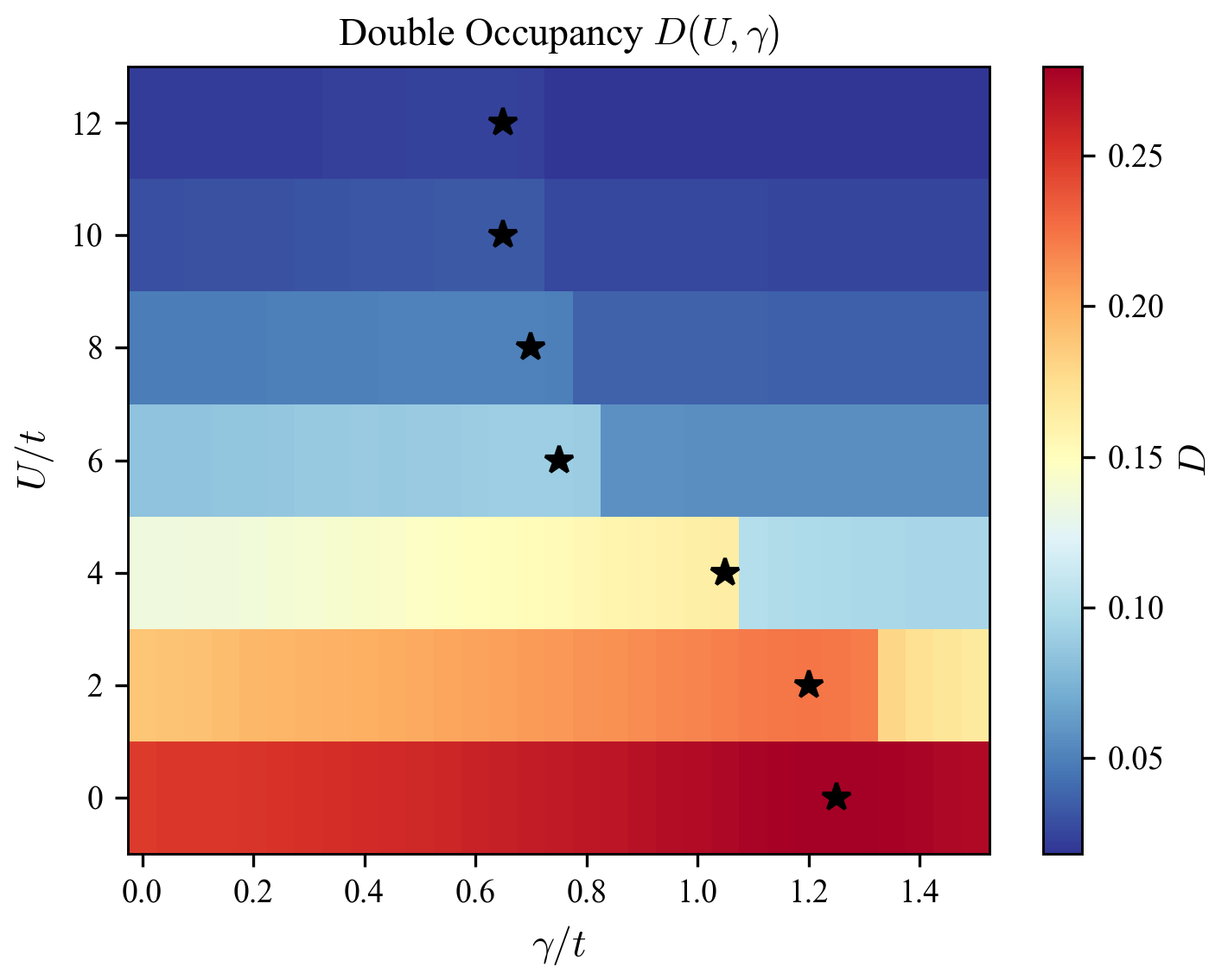}
    \caption{Double occupancy $D(U,\gamma)$ on the $3\times3$ triangular cluster. The golden window appears as a non-monotonic ridge in $\gamma$ whose strength peaks near $U=4t$.}
    \label{fig:heatmap}
\end{figure}

The fixed-sector calculation is internally consistent because the Hatano-Nelson term is a number-conserving one-body hopping: $[H_{\mathrm{NH}},N_\uparrow]=[H_{\mathrm{NH}},N_\downarrow]=[H,S_z]=0$, verified numerically for all $2.7\times10^4$ sampled transitions (zero out-of-sector leakage). This distinguishes it from particle-changing Lindblad jumps, which describe the reservoir route. Figure~\ref{fig:heatmap} visualizes the full $(U,\gamma)$ double-occupancy landscape on the $3\times3$ cluster, showing the golden window as a contiguous ridge in the $(U,\gamma)$ plane whose strength peaks near $U=4t$.

\section{Biorthogonal observable check}
\label{app:biorth}

Table~\ref{tab:biorthogonal} compares the biorthogonal and right-eigenstate double occupancies at $U=4t$ on $3\times3$. The left-eigenstate match via eigenvalue proximity fails at $\gamma/t=0.5,1.0$ (near-degeneracies, biorthogonal norm $\to 0$), but where matching succeeds $D_{\mathrm{bi}}/D_R\approx 0.86$--$0.91$, confirming the golden window is not an artifact of the right-eigenstate choice. On the $2\times3$ cluster, $D_R$ and $D_{\mathrm{bi}}$ trace the same dome ($D_{\mathrm{bi}}/D_R\approx 0.89$, peak $+18\%$ at $\gamma^*\approx 0.5$--$0.6\,t$), with a near-degeneracy window $\gamma/t\in[0.22,0.50]$ where $D_{\mathrm{bi}}$ is undefined.

\begin{table}[!ht]
\centering
\caption{Biorthogonal norm and $D_{\mathrm{bi}}/D_R$ vs.\ $\gamma$ at $U=4t$ ($3\times3$).}
\begin{tabular}{cccc}
\hline
$\gamma/t$ & $|\braket{\psi_L}{\psi_R}|$ & $D_R$ & $D_{\mathrm{bi}}/D_R$ \\
\hline
0.0 & 1.000 & 0.1350 & 1.000 \\
0.3 & 0.424 & 0.1406 & 0.909 \\
0.5 & $\sim 0$ & 0.1461 & divergent \\
0.7 & 0.544 & 0.1521 & 0.874 \\
1.0 & $\sim 0$ & 0.1625 & divergent \\
1.2 & 0.695 & 0.0982 & 0.862 \\
1.5 & 0.800 & 0.0948 & 0.911 \\
\hline
\end{tabular}
\label{tab:biorthogonal}
\end{table}

\section{Exceptional-point analysis}
\label{app:ep}

\begin{figure}[!ht]
    \centering
    \includegraphics[width=0.95\columnwidth]{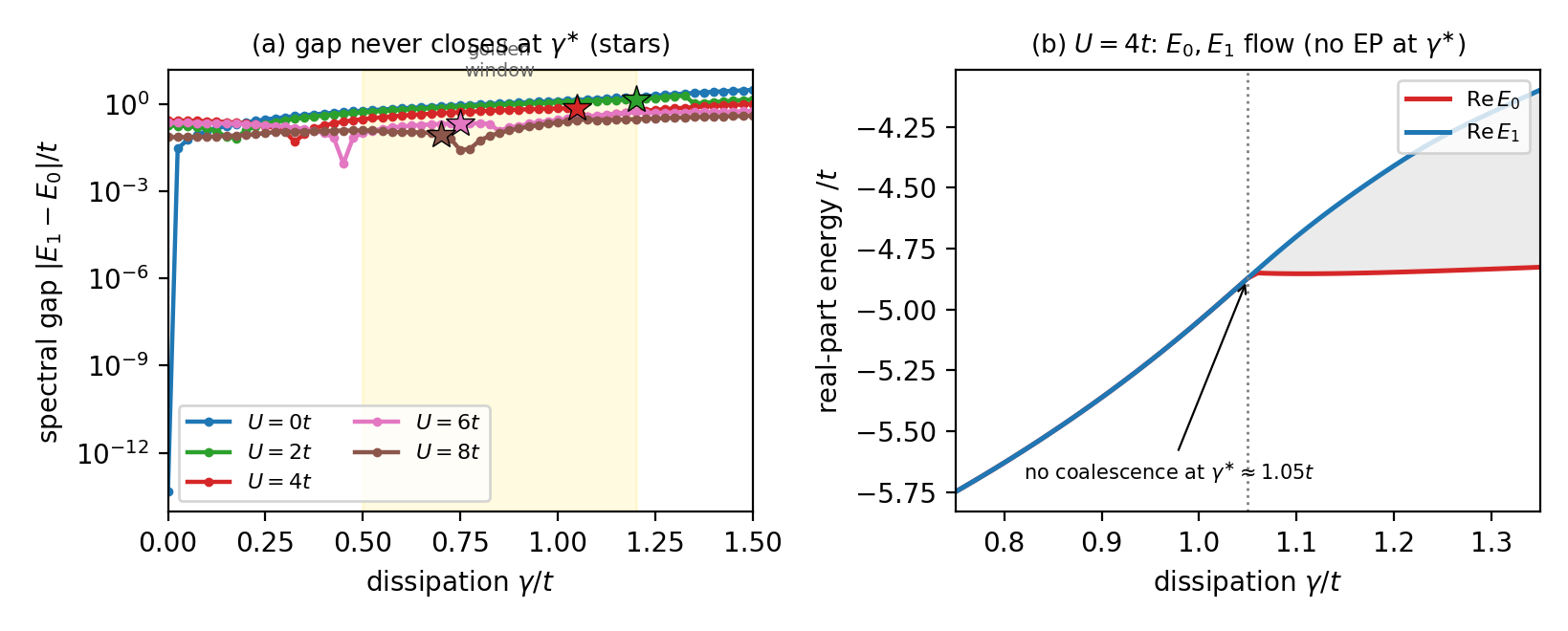}
    \caption{(a) Spectral gap $|E_1-E_0|(\gamma)$ (log scale) for $U=0,2,4,6,8t$; stars mark $\gamma^*(U)$, where the gap is $3$--$24\times$ the global minimum (shaded band: golden window). (b) Real-part flow of $E_0,E_1$ at $U=4t$ on a fine grid around $\gamma^*\approx1.05\,t$ (dotted): no coalescence occurs at the golden window.}
    \label{fig:ep_gap}
\end{figure}

To confirm that the golden window is not tied to an exceptional point (EP), we track the spectral gap $|E_1-E_0|(\gamma)$ on a fine grid ($\Delta\gamma=0.025\,t$) for several couplings [Fig.~\ref{fig:ep_gap}]. The gap never closes inside the golden-window band $\gamma\in[0.5,1.2]\,t$ and is $3$--$24\times$ larger than its global minimum, which sits below $\gamma^*$ at every coupling. The eigenvalue flow is smooth and non-degenerate throughout, confirming smooth NHSE physics rather than a non-analytic spectral singularity.

\section{Open-system consistency: full Lindblad master equation}
\label{app:lindblad}

The pseudo-Hermitian $H_{\rm eff}$ is the controlled description of the \emph{coherent} (Floquet) route, where a high-frequency drive generates the Hatano-Nelson form without injecting quasiparticles~\cite{Yoshida2024}, so the system stays in a dressed ground state and the golden window is an exact zeroth-order prediction. To probe the more dissipative \emph{gain/loss} route we solve the full Lindblad master equation on the $2\times2$ cluster (full Fock space, $2^8=256$ states) with site-resolved gain/loss jumps $L_i^+=\sqrt{g_i}\,c_i^\dagger$, $L_i^-=\sqrt{l_i}\,c_i$ whose imbalance grows along $\hat{x}$ ($g_i+l_i=2g_0$). The non-equilibrium steady state (NESS) reproduces both robust features of the skin mechanism while washing out the non-monotonic dome: (i) the steady-state density piles up at the high-gain boundary (right column $\sim0.92$ vs left $\sim0.71$ at $\gamma=1.5\,t$), exactly the skin-accumulation direction; (ii) the steady-state $D(\gamma)$ grows \emph{monotonically} ($+97\%$), with no downturn, because the over-localization collapse is a $T=0$ ground-state quantum effect smeared at the NESS's finite effective temperature. This separation is itself a prediction: a coherent drive should reveal the full dome including the downturn, whereas dissipative coupling to a metallic layer should show monotonic enhancement.

\section{BCS mean-field analysis}
\label{app:bcs}

The boundary LDOS enhancement $\rho_{\mathrm{skin}}/\rho_0 \sim (2\lambda_{\mathrm{skin}}/L)\,e^{L/\lambda_{\mathrm{skin}}}$ follows from the exponential skin accumulation (prefactor = fraction of skin-localized sites). In a BCS mean-field treatment the NHSE-enhanced Fermi-level DOS $\rho_{\mathrm{eff}}(\gamma)$ rises for $\gamma<\gamma^*$ (LDOS amplification) then falls for $\gamma>\gamma^*$ (spectral collapse from over-localization), producing a non-monotonic $T_c(\gamma)$ dome. At the peak ($U=4t$) the BCS-mapped ratio is $T_c^{\max}/T_c^0\sim 1$--$1.5\times$, mirroring the $\sim21\%$ $D$ enhancement; this is a qualitative re-expression of the same boundary-LDOS signal already captured by $D(\gamma)$, not an independent prediction of $T_c$. Mapping to physical units ($t_{\mathrm{moire}}\sim5$ meV) gives $T_c^{\max}\sim 0.04\,t\sim 2$ K, consistent with tWSe$_2$~\cite{Xia2025}.

\section{Experimental parameter mapping}
\label{app:exp_params}

\begin{table}[!ht]
\centering
\caption{Parameter mapping for moir\'{e} experimental platforms.}
\begin{tabular}{lccc}
\hline
Platform & $t_{\mathrm{moire}}$ & $U/t$ & $\gamma^*/t$ \\
\hline
tWSe$_2$ & 5--10 meV & 4--8 & 0.7--1.1 \\
tMoTe$_2$ & 3--8 meV & 6--10 & 0.6--0.8 \\
MATBG & 1--3 meV & 8--12 & 0.65--0.7 \\
\hline
\end{tabular}
\label{tab:exp_params}
\end{table}

For tWSe$_2$ with $t\sim5$ meV~\cite{Xia2025,Wu2019}, $\gamma^*\sim1.05\,t\sim5$ meV corresponds to a non-reciprocal resistance asymmetry $\Delta R/R\sim\gamma/t\sim1$. The 2D NHSE has been realized in an ultracold-atom Fermi gas~\cite{Zhao2025}; current solid-state devices probe the onset regime ($\gamma/t\sim0.1$, $\Delta R/R\sim10\%$). The key signature---$T_c$ first increasing then decreasing with gate-controlled dissipation---requires only qualitative observation.

\begin{acknowledgments}
This work was supported by the Scientific Research Project (No.WU2025B011) and the Start-up Funding of Westlake University. The authors used AI-assisted language editing during the preparation of this manuscript and take full responsibility for the final content.
\end{acknowledgments}

\textit{Data Availability}---The exact diagonalization code and numerical data are available from the corresponding author upon reasonable request.

\bibliographystyle{apsrev4-2}
\bibliography{references}

\end{document}